\documentclass[12pt,tightenlines,eqsecnum,floats,shownopacs,nofootinbib,amsmath,amssymb,aps,prd]{revtex4}

\usepackage{color}
\usepackage{graphicx}
\usepackage{amsmath,amssymb}

\newcommand{\be}{\begin{equation}}
\newcommand{\ee}{\end{equation}}
\newcommand{\bea}{\begin{eqnarray}}
\newcommand{\eea}{\end{eqnarray}}

\def\x{\vec{x}}
\def\h{\hat}
\def\T{\underbar{T}}
\def\dd{{\rm d}}
\def\G{G_{\rm N}}
\usepackage[colorinlistoftodos]{todonotes}
\usepackage{epstopdf}

\begin{document}

\title{Preferred instantaneous vacuum for linear scalar fields in cosmological space-times}

\author{Ivan Agullo$^{1,2}$}
\email{agullo@lsu.edu}
\author{William Nelson$^{2}$}
\email{nelson@gravity.psu.edu}
\author{Abhay Ashtekar$^{2}$}
\email{ashtekar@gravity.psu.edu}
\affiliation{$^{1}$ Department of Physics and Astronomy, Louisiana State University, Baton Rouge, LA 70803, U.S.A.,}
\affiliation{$^{2}$ Institute for
Gravitation and the Cosmos \& Physics
  Department, Penn State, University Park, PA 16802, U.S.A.}

\begin{abstract}

We discuss the problem of defining a preferred vacuum state at a given time for a quantized scalar field in Friedmann, Lema\^itre, Robertson, Walker (FLRW) space-time.  Among the infinitely many homogeneous, isotropic vacua available in the theory, we show that there exists at most one for which every Fourier mode makes a vanishing contribution to the adiabatically renormalized  energy-momentum tensor at any given instant. For massive fields such a state exists in the most commonly used backgrounds in cosmology and,  within the adiabatic regularization scheme, provides a natural candidate for the ``ground state''  at that instant of time. The extension to the massless  and the conformally coupled case are also discussed.

\end{abstract}

\pacs{04.62.+v,  98.80.-k}
%\pacs{04.60.Kz, 04.60.Pp, 98.80.Qc}

\maketitle

\section{Introduction}
\label{s1}

Perhaps the most important lesson we have learned from quantum field theory in arbitrarily  curved space-times is the absence of a preferred  vacuum state  \cite{birrell-davies,sf-book,Waldbook,parker-book}.  Interesting phenomena such as particle creation in an expanding universe \cite{ParkerThesis,parker69}, the Hawking effect in black hole backgrounds \cite{hawking}, and the Unruh effect in Minkowski space-time \cite{unruh} rely on this fact. In highly symmetric space-times like Minkowski or de Sitter space, the underlying isometries are a powerful tool to single out preferred vacua. One proceeds by requiring two conditions: i) The vacuum must be invariant under the full  group of symmetries of the background metric; ii) The vacuum must be ultraviolet (UV) regular, namely the short-distance or large-frequency structure of the state must approach the behavior found in flat space {\em at an appropriate rate}. The adiabatic regularity condition in homogeneous space-times or the Hadamard condition in more generic backgrounds are concrete ways of implementing the second requirement \cite{Waldbook,parker-book}. UV regularity, among other things, guarantees that  composite operators such as the energy-momentum tensor can be satisfactorily  renormalized in the Hilbert space of physical states.  In Minkowski and  de Sitter space-times these two requirements are indeed strong enough to {\em uniquely} single out a vacuum state, the so-called Minkowski and Bunch-Davies vacuum, respectively. %just moved this sentence up for better continuity.

However, for less symmetric backgrounds one finds infinitely many such states. In particular, this is the case for the Friedmann, Lema\^itre, Robertson, Walker (FLRW) space-times with line element $\dd s^2=a(\eta)^2\, (-\dd\eta^2+\dd\vec{x}^2)$ in the conformal time $\eta$. Although they are not maximally symmetric, these space-times carry three space translations and three rotations, significantly simplifying the analysis. Because of the central importance of these space-times in cosmology, repeated attempts have been made to select preferred vacua for test quantum fields using these simplifications. However, to our knowledge, a satisfactory solution has not emerged. Perhaps the simplest idea, that appears compelling at first, is to try to define the instantaneous vacuum as the ground state of the Hamiltonian operator at that instant of time. However, as shown in \cite{fulling78}, this strategy faces two key difficulties. First, to define the Hamiltonian, one has to make a choice of canonical variables and this freedom introduces an ambiguity in the choice of states. Second, even after making a specific choice, the resulting state fails to have the desired UV regularity, except in very specific situations.

The goal of this paper is to propose an alternate strategy which is motivated by the same physical considerations but which is free of the two limitations. Specifically, we avoid the ambiguities associated with the choice of canonical variables by working only with \emph{space-time fields} and, from the start, we restrict ourselves to states that are UV regular. In essence, the key idea is to select the instantaneous vacuum $|0\rangle$ at $\eta=\eta_{0}$ by demanding that the expectation value of the stress-energy tensor $\h{T}_{ab}(\x, \eta_{0})$ in $|0\rangle$ should vanish: $\langle 0|\h{T}_{ab}(\vec{x},\eta_0)|0\rangle=0$. 

Because it is constructed directly from space-time fields, without reference to canonically conjugate variables,  the energy-momentum tensor is free of the ambiguities appearing in the  Hamiltonian. Furthermore, the expectation value $\langle 0|\h{T}_{ab}(\vec{x},\eta_0)|0\rangle$ has fundamental physical significance because it is the vehicle through which matter fields source gravity in the semiclassical approximation, and its conservation law provides rich information already in the test field approximation.
However, the formal expression of this vacuum expectation value (VEV) is UV divergent and requires renormalization even in a noninteracting theory. (This is the generalization of the Minkowskian normal ordering procedure to curved space-times.) In this paper we will use the adiabatic renormalization \cite{ParkerThesis,parker69, parker-fulling74} which is particularly transparent for computations in homogeneous space-times. This scheme takes advantage of the translational symmetry of the background and renormalizes the energy-momentum tensor by subtracting suitable counter-terms {\em using Fourier modes}. More precisely, the stress-energy tensor $\h{T}_{ab}(\x,\eta)$ is a composite operator. By expanding each field operator $\hat{\phi} (\x, \eta)$ in its (formal) expression in terms of its Fourier modes, one can express the expectation value 
of $\h{T}_{ab}(\x,\eta)$ as an integral in the momentum space:
\be \langle 0|\h{T}_{ab}(\vec{x},\eta_0)|0\rangle = \int {\dd^3k} \, \underbar{T}_{ab} (\vec{k},\eta_{0}) \, ,\ee 
As explained in section \ref{s2}, the renormalized expression is given by subtracting the appropriate counter-term $C_{ab}(\vec{k},\eta)$ \emph{for each} $\vec{k}$:
\be \langle 0|\h{T}_{ab}(\vec{x},\eta_0)|0\rangle_{\rm ren}=\int {\dd^3k} \, [\underbar{T}_{ab}(\vec{k},\eta) - C_{ab}(\vec{k},\eta)] \, ,\ee
We show that, whenever it is possible to find a homogeneous and isotropic state that satisfies $\T_{ab}(\vec{k},\eta_0)=C_{ab}(\vec{k},\eta_0)$ for every $\vec{k}$, that state is {\em unique} and  UV-regular. In particular, in this state $\langle 0|\h{T}_{ab}(\vec{x}, \eta_0)|0\rangle_{\rm ren}=0$. Note however that, because $\T_{ab}(\vec{k},\eta_0)-C_{ab}(\vec{k},\eta_0)$ is not necessarily positive, our requirement that it vanish for each $\vec{k}$ is stronger than simply asking $\langle 0|\h{T}_{ab}(\vec{x}, \eta_0)|0\rangle_{\rm ren}=0$: We are excluding the possibility of a cancelation between contributions from different $\vec{k}$-modes. 

The resulting state is tailored to the time $\eta_0$ because, generically,  $\langle 0|\h{T}_{ab}(\vec{x}, \eta)|0\rangle_{\rm ren}$ will be nonzero at any other time $\eta$. For this reason we will call it the preferred \emph{instantaneous vacuum}, and  
denote it by $|0_{\eta_0}\rangle$, where $\eta_{0}$ is the instant to which it refers.  Renormalization of the energy-momentum tensor can then be understood as ``a {\em time-dependent} normal ordering'' with respect to  the  $\eta$-family of preferred instantaneous vacua. Furthermore, the expectation value $\langle 0_{\eta_0}|\h{T}_{ab}(\vec{x},\eta_1)|0_{\eta_0}\rangle_{\rm ren}$ at another time $\eta_1 >\eta_{0}$ can be interpreted as the energy-momentum transferred to the scalar field by the dynamical background geometry. These features serve to bring out the physical meaning of our instantaneous vacuum. Note, however, that the notion depends  
on our choice of adiabatic renormalization because of the mode by mode subtraction involved. One can imagine using a variation of this strategy tailored to another renormalization scheme and the resulting strategy may well yield a different notion of an instantaneous vacuum. However, one does need renormalization to speak of composite operators such as energy and momentum density and it is non-trivial that there exists a scheme which enables one to select instantaneous vacua in a large number of physically important cosmological situations.

The approach presented here has some similarities with a part of the analysis carried out out in \cite{Anderson-Molina-Paris-Mottola}. In that work, among many other interesting results, adiabatic renormalization was used to obtain a preferred definition of particles at a given time, and the resulting definition was applied to discuss the  creation of particles by the expansion of the universe. In the present paper we discuss states that make the VEV of the full renormalized energy-momentum tensor vanish. No particle interpretation of the field theory is required here. 

We will conclude this introduction with a couple of conceptual remarks. 
An important tenet of quantum field theory in curved backgrounds is that the renormalization procedure can only make use of the {\em local} properties of the space-time geometry, namely curvature tensors and its derivatives at a point \cite{Waldbook}. But our condition $\langle 0_{\eta_0}|\h{T}_{ab}(\vec{x},\eta_0)|0_{\eta_0}\rangle_{\rm ren}=0$ is global in space because it is required to hold for all $\x$. However, because we require the states to be spatially homogeneous, satisfaction of this condition at one $\x$ implies that it holds for all $\x$ at $\eta=\eta_{0}$. Thus, while states are `global notions', in our strategy the spatial aspect of this global character is ensured by asking that the state $|0_{\eta_{0}}\rangle$ be spatially homogeneous. 

The second point concerns the \emph{existence} of the preferred instantaneous vacuum. General arguments indicate that it cannot exist for arbitrary values of the mass $m$ and coupling to the curvature $\xi$. For instance, it is well-known that for a conformally coupled scalar field ($m=0$ and $\xi=\frac{1}{6}$) the trace of the renormalized energy-momentum tensor is {\em nonzero} and {\em independent} of the quantum state of the field. This is the well known trace anomaly \cite{capper-duff,desser-duff-isham}. As one would expect, our strategy fails to select a state in these cases (see section \ref{s4}). Thus, the strategy succeeds in selecting a preferred state in generic physically interesting situations, neatly bypassing the special cases in which conceptual obstacles are already known to exist.

The plan of the paper is the following. We work in a spatially flat FLRW space-time (although we do not envisage significant difficulties in extending the analysis to other homogeneous space-times). In section \ref{s2} we summarize the expression for the renormalized energy-momentum tensor.  In section \ref{s3} we analyze the minimally coupled, massive scalar field and provide a criterion for the existence of the preferred instantaneous vacuum. We check that the criterion is met in the space-times commonly considered in cosmology, including radiation-dominated, matter-dominated, FLRW space-times (except very near the big bang singularity), in de Sitter space, and of course, Minkowski space. In section \ref{s4} we consider the massless, minimally coupled case and  show that the preferred vacuum does not exist. However, the problem is  similar to the one encountered in de Sitter space in the massless limit of the Bunch-Davies vacuum \cite{allen} and one can work around it  in the same fashion \cite{allen-folacci}. In section \ref{s5} the conformally coupled case is discussed, and it is shown that it is {\it not} possible to find a state with zero expectation value of the energy-momentum tensor, unless $a(\eta)$ is very special ({\it e.g.} constant). Section \ref{s6}  provides a summary, a discussion of an interpretation of the instantaneous vacuum in the framework of semiclassical gravity, and some final comments.

Our conventions:  signature is $-+++$; curvature tensors are defined as: $R_{abc}{}^{d} v_{d} = 2 \nabla_{[a} \nabla_{b]}\, v_{c};\,\, R_{ac} = R_{abc}{}^{b};\,\, R = g^{ac}R_{ac}$;\,\, and $c=\hbar =1$.

\section{Qft in $K=0$ FLRW backgrounds and renormalized VEV of the energy-momentum tensor \label{s2}}

In this section we summarize the expression for the renormalized VEV of the energy-momentum tensor in adiabatic regularization in spatially flat FLRW. For more details see \cite{parker-book, birrell-davies, sf-book}. Consider a noninteracting, real scalar field satisfying the Klein-Gordon equation $(\Box -m^2-\xi \, R)\hat \phi(\vec{x},\eta)=0$, where $R=6 {a''}/{a^3}$ is the scalar  curvature  of the FLRW metric $ds^2=a(\eta)^2\, (-d\eta^2+d\vec{x}^2)$, and prime denotes the derivative with respect to conformal time $\eta$. We analyze here the minimally coupled case $\xi=0$ and $m\not=0$, leaving the massless minimally coupled case  and conformally coupled case ($\xi=1/6$) for sections \ref{s4} and \ref{s5}, respectively. 

The underlying homogeneity can be used to Fourier expand the field operator and represented it as

\be \label{fieldop}\hat \phi(\vec{x},\eta)= \frac{1}{(2\pi)^3}\int \dd^3k \,  [\hat A_{\vec{k} } \, \varphi_{\vec{k}}(\eta)+\hat A^{\dagger}_{-\vec{k}} \, \bar \varphi_{-\vec{k}}(\eta)]\,  e^{i\vec{k}\cdot\vec{x}}\,  ,\ee
where `bar' denotes complex conjugation. The basis functions $\varphi_{\vec{k}}(\eta)$ are solutions of the wave equation 

\be \label{modeq} \varphi''_{\vec{k}}(\eta)+2\frac{a'}{a} \varphi'_{\vec{k}}(\eta)+(k^2+m^2 a^2)\varphi_{\vec{k}}(\eta)=0 \, , \ee
and if they are chosen to satisfy the normalization conditions $ \varphi_{\vec{k}}\bar \varphi'_{\vec{k}}-\varphi'_{\vec{k}}\bar \varphi_{\vec{k}}=i\, a^{-2}$ and  $ \varphi_{\vec{k}} \varphi'_{-\vec{k}}-\varphi'_{\vec{k}}\varphi_{-\vec{k}}=0$ at some instant of time, then the time-independent operators $\hat A_{\vec{k} }$ and $\hat A^{\dagger}_{\vec{k} }$ satisfy the algebra of creation and annihilation operators: $[\hat A_{\vec{k}},\hat A^{\dagger}_{\vec{k}' }]= (2\pi)^3 \delta^3(\vec{k}-\vec{k}')$, $[\hat A_{\vec{k}}\, ,\hat A_{\vec{k}' }]=[\hat A^{\dagger}_{\vec{k}}\, ,\hat A^{\dagger}_{\vec{k}' }]=0$. One then defines the vacuum $|0\rangle$ as the state annihilated by all $\hat A_{\vec{k}}$, and generates the Fock space by repeatedly acting on it with creation operators. 

The vacuum defined in this way is tailored to the definition of  the operators $\hat A_{\vec{k}}$. In turn, these operators are uniquely determined by the specification of a complete set of mode functions $\varphi_{\vec{k}}(\eta)$ for all $\vec{k}$: Eq. (\ref{fieldop}) and the normalization condition imply $\hat A_{\vec{k}}%:=\hat A[\varphi_{\vec{k}}] (Abhay:  This new symbol is unnecessary and not used again.)
=-i\int \dd^3 x \, [\hat \phi(\vec{x},\eta) \, {\buildrel\leftrightarrow\over{\partial}}_{\eta} \,  \bar \varphi_{\vec{k}}(\eta)\, e^{-i\vec{k}\cdot \vec{x}}]$.
Therefore, a complete family of normalized solutions $\varphi_{\vec{k}}(\eta)$ to equation (\ref{modeq}) determines a vacuum. But the correspondence is not one to one. The sets $\{\varphi_{\vec{k}}(\eta)\}$ and $\{\varphi_{\vec{k}}(\eta)\, e^{i \theta_{\vec{k}}}\}$  that only differ by a time-independent phase factor determine the same vacuum.%
\footnote{As explained in section V.B of \cite{aan2}, there is a 1-1 correspondence between these equivalence classes of basis and complex structures $J$ on the space $\mathcal{S}$ of real, classical solutions to the field equations, which are compatible with the natural symplectic structure $\Omega$ on $\mathcal{S}$ in the sense that $(\mathcal{S}, \Omega, J)$ is a K\"ahler space.}

The resulting vacua are all translational invariant, but we can impose an additional condition on the mode functions $\varphi_{\vec{k}}(\eta)$ to ensure that they are also rotationally symmetric. This is achieved by demanding that mode functions depend only on the norm of the wave vector $k=|\vec{k}|$, rather than on its three independent components. As is well known, the rotational invariance can be demonstrated by writing down the associated two-point function
\be \label{twopoint} \langle 0|\hat\phi(\vec{x}_1,\eta_1)\hat\phi(\vec{x}_2,\eta_2)|0\rangle =\int \frac{\dd^3k}{(2\pi)^3} \, \varphi_k(\eta_1)\bar \varphi_k(\eta_2)\ e^{i \vec{k}\cdot(\vec{x}_1-\vec{x}_2)} \, , \ee
which by inspection displays invariance under these symmetries.  

To summarize, in FLRW there is a one-to-one correspondence between equivalence classes of families of solutions $\{ \varphi_k (\eta)\}$ which differ by time-independent phase factors $e^{i\theta_{\vec{k}}}$ and translationally and rotationally invariant vacuum states. \\

The classical expression for the energy-momentum tensor of a minimally coupled scalar field is 
\be T_{ab}= \nabla_a\phi\nabla_b\phi-\frac{1}{2}g_{ab}g^{cd} \nabla_c\phi\nabla_d\phi -\frac{1}{2}m^2g_{ab}\phi^2 \, .\ee
In the quantum theory, the expectation value of the operator $\h{T}_{ab}$ in a homogeneous and isotropic vacuum state takes the perfect fluid form
\be \label{pf} \langle 0|\h{T}_{ab}|0\rangle =g_{ab} \langle \h{p} \rangle +(\langle \h{p} \rangle +\langle \h{\rho} \rangle )\, u_a u_b \, ,\ee 
where $u^a$ is the unit vector normal to the homogeneous and isotropic hypersurfaces. In terms of the modes $\varphi_k(\eta)$ defining the vacuum, the \emph{formal} expressions of the expectation values of energy density and pressure are
\be \label{rho1} \langle \h{\rho} \rangle:= \frac{1}{(2\pi)^3} \int \dd^3k \, \rho[\varphi_k]=\frac{1}{(2\pi)^3} \int \dd^3k\,  \frac{1}{2 a^2}\,  \big(|\varphi_k'|^2+w^2 |\varphi_k|^2\big)\, , \ee
\be\label{p1}  \langle \h{p} \rangle:= \frac{1}{(2\pi)^3} \int \dd^3k \, p[\varphi_k] =\frac{1}{(2\pi)^3} \int  \dd^3k \, \frac{1}{2 a^2}\,   \big(|\varphi_k'|^2-\frac{1}{3} (w^2+ 2 m^2) |\varphi_k|^2\big)\, ,\ee
where the time-dependent frequency $w(\eta)$ is given as usual by 
\be w(\eta)=\sqrt{k^2+m^2 a(\eta)^2}\,. \ee
The  VEV of the trace of the  energy-momentum tensor  is $\langle \h{T} \rangle= 3\, \langle \h{p} \rangle -\langle \h{\rho} \rangle$. These expressions for the components of $\langle \h{T}_{ab}\rangle $ are only  formal because they  diverge in the $k\to\infty$ limit as $k^4$, regardless of the form of $a(\eta)$. Regularization and renormalization are required to extract the finite, physically relevant result. As mentioned in the Introduction, in this paper we use the adiabatic renormalization method developed by Parker and Fulling in \cite{ParkerThesis,parker69, parker-fulling74}. More recent accounts can be found in \cite{parker-book,sf-book} and a succinct summary, most closely related to our present discussion, is given in section IV of \cite{aan2}. This method removes the UV divergences by subtracting the adiabatic counterterms mode by mode, under the $\vec k$-integral: 
\be \label{rho} \langle \h{\rho} \rangle_{\rm ren}= \frac{1}{(2\pi)^3} \int \dd^3k \, \big(\rho[\varphi_k] -C_{\rho}(\eta,k,m)\big)\, , \ee
and,
\be \label{p} \langle \h{p} \rangle_{\rm ren}=  \frac{1}{(2\pi)^3} \int \dd^3k \, \big(p[\varphi_k] -C_{p}(\eta,k,m)\big)\, .\ee
Here $C_{\rho}(\eta,k,m)$ and $C_{p}(\eta,k,m)$ are the  rather long expressions (\ref{Crhoxi0}) and (\ref{Cpxi0}), given in the Appendix (see also \cite{anderson-parker}). They are independent of the state in which the expectation values are evaluated, and therefore independent of the choice of modes $\varphi_k$. They are functions of $k$, constructed entirely from the scale factor $a(\eta)$ and its four first time derivatives at time $\eta$. Therefore, to renormalize the stress energy tensor using adiabatic renormalization, $a(\eta)$ has to be a $C^{4}$ function.% 
\footnote{The stress energy tensor is a composite operator of dimension 4. More generally, to regularize and renormalize an operator product of dimension $n$ one needs adiabatic regularity of order $n$ which requires $a(\eta)$ to be $C^{n}$.} 

As is well known, \cite{birrell-davies, sf-book,Waldbook,parker-book}, in order to have a physically satisfactory quantum field theory one needs to impose restrictions on the allowed quantum states. These are the regularity conditions mentioned in section \ref{s1}. In the adiabatic approach one restricts physical states to be of 4th adiabatic order. This requirement is implemented by demanding asymptotic conditions on the family of solutions $\varphi_k(\eta)$ defining the vacuum, in the limit  $w\to \infty$. One requires the modes $\varphi_k(\eta)$ to approach Minkowski positive frequency solutions ($e^{-iw\eta}/\sqrt{2w}$) \emph{at the appropriate rate}, specified by  the following behavior in the $k\to \infty$ limit:
\be \label{adcond} |\varphi_k(\eta)|= | \varphi_k^{(4)}(\eta)|\Big(1+\mathcal{O}(w^{-(4+\epsilon)}) \Big)\,  \hspace{0.25cm} {\rm and}\hspace{0.25cm} |\partial_{\eta}\varphi_k(\eta)|= | \partial_{\eta} \varphi_k^{(4)}(\eta)|\Big(1+\mathcal{O}(w^{-(4+\epsilon)}) \Big)\, , \ee 
with $\epsilon$ a strictly positive real number and 
\be  \varphi_k^{(4)}(\eta)= \frac{1}{a(\eta)\sqrt{2 \, W_k^{(4)}(\eta)}} e^{-i \int^{\eta} W_k^{(4)}(\eta') \dd\eta'} \,\ee 
where $W_k^{(4)}(\eta)=W_0+W_2+W_4$, with
\bea \label{adexp}W_0&=&w\, ; \nonumber\\ W_2&=& \frac{3 a w'^2-4w^2a''-2aww''}{8 a w^3}\, ;\nonumber\\ W_4&=&\frac{1}{128 a^3 w^7} (-297 a^3 w'^4+32 w^4a'^2a''+80aw^3a'w'a''+152a^2w^2w'^2a''-32aw^4a''^2+\nonumber\\&+&396 a^3ww'^2w''-48a^2w^3a''w''-52a^3w^2w''^2-32aw^4 a'a'''-80a^2w^3w'a'''-\nonumber\\&-&80 a^3w^2w'w'''+16a^2w^4a''''+8a^3w^3w'''')\, .\eea
If conditions (\ref{adcond}) are satisfied at some time $\eta_0$, the wave equation (\ref{modeq}) guarantees they are satisfied for all $\eta$. For further details about the adiabatic expansion see \cite{parker-book,sf-book}. Mode functions $\varphi_k(\eta)$ satisfying requirements (\ref{adcond}) are called 4th adiabatic order modes, and the vacuum  $|0_{\eta_{0}}\rangle$ they define is a quantum state of 4th adiabatic order. (Elements of the Hilbert space obtained by acting repeatedly by a finite but arbitrarily large number of creation operators provides a dense subspace of states all of which are of 4th adiabatic order.) Note that (\ref{adcond}) imposes only asymptotic restrictions. Therefore there are infinitely many choices of modes $\{\varphi_{k}\}$ of 4th adiabatic order and hence of vacua $|0\rangle$ of 4th adiabatic order.

{\bf Remark:} In spatially compact space-times, Hilbert spaces constructed from different adiabatic vacua are unitarily equivalent. In this sense, the adiabatic condition selects a unique Hilbert space and the associated representation of the quantum theory. This is not the case if space is noncompact. In that situation inequivalent representations appear, even if states are adiabatic up to all orders (the same happens for Hadamard states \cite{Waldbook}). But this mathematical inequivalence is considered to be physically spurious, since the resulting theories are physically indistinguishable when measurements are restricted to a finite region of space. (See, e.g. section 2.3.2 of \cite{aan3}.)

\section{Preferred instantaneous vacuum for massive, minimally coupled scalar fields}\label{s3}

Let us start only with  the (spatial) translational invariance of the background geometry to perform Fourier transform and incorporate the rotational invariance in a second step. Then, as described in section \ref{s2}, there is a one-to-one correspondence  between equivalence classes of families of normalized solutions to equation (\ref{modeq}),  $\{ \varphi_{\vec{k}}(\eta)\}$, that differ only by a time-independent phase factor, and Fock vacua $|0\rangle$. Since equation (\ref{modeq}) is a second-order O.D.E., the modes $\varphi_{\vec{k}}(\eta)$ are uniquely determined by their initial data $\{\varphi_{\vec{k}}(\eta_0), \,\,\varphi'_{\vec{k}}(\eta_0) \}\, \in \mathbb{C}^{2}$ at any given time $\eta_0$. Once the normalization condition and the irrelevant phase factor are taken into account, two independent real parameters for each  $\vec{k}$ are sufficient  to unambiguously  determine solutions $\varphi_{\vec{k}}(\eta)$. They can be conveniently chosen as $\Omega_{\vec{k}}(\eta_0)$ and $V_{\vec{k}}(\eta_0)\, \in \mathbb{R}$ satisfying $\Omega_{-\vec{k}}(\eta_0)=\Omega_{\vec{k}}(\eta_0)$ and $V_{-\vec{k}}(\eta_0)=V_{\vec{k}}(\eta_0)$. In terms of these parameters we can set%
\footnote{This parametrization for initial data was already used in \cite{Anderson-Molina-Paris-Mottola}.}
\be \label{initial-data} \varphi_{\vec{k}}(\eta_0)= \frac{1}{a(\eta_0) \sqrt{2 \, \Omega_{\vec{k}}(\eta_0)}} \, ; \hspace{1cm}  \varphi'_{\vec{k}}(\eta_0)= \left(-i\, \Omega_{\vec{k}}(\eta_0)+\frac{V_{\vec{k}}(\eta_0)}{2}-\frac{a'(\eta_0)}{a(\eta_0)}\right) \, \varphi_{\vec{k}}(\eta_0) \, .\ee
Therefore, a set of such real numbers $\Omega_{\vec{k}}(\eta_0)$ and $V_{\vec{k}}(\eta_0)$ for every $\vec{k}$ is in one-to-one correspondence with the set of homogeneous Fock vacua. We will take advantage of this correspondence to find the desired preferred instantaneous vacuum $|0_{\eta_0}\rangle$ at time $\eta_0$. As described in section \ref{s1}, the strategy is to look for states  $|0_{\eta_0}\rangle$ satisfying three requirements:

\begin{enumerate}

\item \emph{The symmetry requirement:}\, $|0_{\eta_0}\rangle$ shares the symmetries of the background metric, namely (spatial) translational \emph{and} rotational invariance. As discussed in section \ref{s2}, this is guaranteed if the solutions $\varphi_{\vec{k}}(\eta)$  depend only on the norm $k$ of the wave vector $\vec{k}$. This is the case if and only if $\Omega_{\vec{k}}(\eta_0)=\Omega_{k}(\eta_0)$ and $V_{\vec{k}}(\eta_0)=V_{k}(\eta_0)$ for all $\vec{k}$.
 
\item \emph{The regularity requirement:}\, $|0_{\eta_0}\rangle$ is a quantum state of 4th adiabatic order. This will be the case if and only if $\varphi_k(\eta_0)$ and $\varphi'_k(\eta_0)$ satisfy (\ref{adcond}) -- (\ref{adexp}). This, in turn, is guaranteed if and only if $\Omega_{k}(\eta_0)$ and $V_{k}(\eta_0)$ satisfy the following asymptotic conditions as $w\to\infty$:
$$\Omega_{k}(\eta_0)=W_k^{(4)}(\eta_0)+\mathcal{O}(w^{-(4+\epsilon)}) \, ; \qquad   V_{k}(\eta_0)=\frac{\partial_{\eta}W_k^{(4)}}{W_k^{(4)}}\Big  |_{\eta_0}+\mathcal{O}(w^{-(4+\epsilon)}) \, ,$$ 
where $W_k^{(4)}(\eta)$ is defined by  (\ref{adexp}) and $\epsilon >0$.
 
\item \emph{The `instantaneous vacuum' requirement:}  For each $\vec{k}$ we require $\rho[\varphi_{k} (\eta_{0})] - C_{\rho}(\eta, k, m) =0$ and $p[\varphi_{k} (\eta_{0})] - C_{p}(\eta, k, m) =0$ so that the renormalized expectation value $\langle 0_{\eta_0}|\hat T_{ab}(\eta_0) |0_{\eta_0}\rangle_{\rm ren}$ of the stress tensor vanishes identically, mode by mode.
\end{enumerate}
At first these requirements appear to impose an overconstrained set of conditions on $\Omega_{k}(\eta_0)$ and $V_{k}(\eta_0)$. Therefore, there is no {\em a priori} guarantee that a solution would exist. We now investigate existence and uniqueness. 

The third condition requires $\rho[\varphi_k(\eta_0)] =C_{\rho}(\eta_0,k,m)$ and $p[\varphi_k(\eta_0)] =C_{p}(\eta_0,k,m)$ for all $\vec{k}$, where $\rho[\varphi_k] $ and $p[\varphi_k] $ were defined in equations (\ref{rho1}) and (\ref{p1}), and $\varphi_k(\eta_0)$ is given by (\ref{initial-data}). This is a quadratic system of algebraic equations for $\Omega_{k}(\eta_0)$ and $V_{k}(\eta_0)$. The solutions are
\be \label{omega} \Omega_{k}(\eta_0)=-\frac{2 \, w^{2}(\eta_0) + m^2 \, a^{2}(\eta_0)}{6 \, a^{4}(\eta_0)\,( C_p(\eta_0,k,m)- \,  C_{\rho}(\eta_0,k,m))} \, ; \ee 
\be \label{v} V^{(\pm)}_{k}(\eta_0)=2\frac{a'(\eta_0)}{a(\eta_0)}\mp 2 \sqrt{-w^2(\eta_0)+4\, a^{4}(\eta_0) \, C_{\rho}(\eta_0,k,m)\Omega_{k}(\eta_0)-\Omega_{k}^2(\eta_0)} \, .\ee
Additionally, $\Omega_{k}(\eta_0)$ must be positive and both $\Omega_{k}(\eta_0)$ and $V_{k}(\eta_0)$ must be finite and real for the  initial data (\ref{initial-data}) to define normalized solutions. These requirements translate to the following conditions 
\be \label{cond1} \infty > \Omega_{k}(\eta_0)>0\, , \ee 
\be \label{cond2} \infty > r_k(\eta):=  -w^2(\eta_0)+4\, a^{4}(\eta_0)\, C_{\rho}(\eta_0,k,m)\Omega_{k}(\eta_0)-\Omega_{k}^2(\eta_0)\ge 0 \, .\ee
If they are satisfied, then $ \Omega_{k}(\eta_0)$ and $V^{(\pm)}_{k}(\eta_0)$ define vacuum states satisfying conditions i) and iii) in our list.

But would the resulting vacua meet the regularity condition ii), {\it i.e.}, are they states of 4th adiabatic order? Note that the only remaining freedom is the choice of sign in (\ref{v}). If $(\Omega_{k}(\eta_{0}),\, r_{k}(\eta_{0}))$ satisfy (\ref{cond1}) and (\ref{cond2}), a detailed examination shows that the vacuum state constructed from $\Omega_{k}(\eta_0)$ and $V^{(+)}_{k}(\eta_0)$ is of 4th adiabatic order in an {\em expanding} universe ($a'(\eta)\ge 0$), while $ \Omega_{k}(\eta_0)$ and $V^{(-)}_{k}(\eta_0)$ provide the satisfactory solution in the {\em contracting} case. {\em Therefore if the solution $|0_{\eta_{0}}\rangle$ exists, then it is unique.}

To summarize, equations (\ref{cond1}) and (\ref{cond2}) provide the necessary and sufficient conditions for the existence of a Fock vacuum satisfying our three requirements. Assuming existence of such an $\Omega_{k}(\eta_0)$, using the expression (\ref{twopoint}) of the two-point function, the resulting state $|0_{\eta_0}\rangle$ can be shown to be regular both in  the infrared limit $k\to 0$ as well as in the UV limit $k\to \infty$ at $\eta=\eta_0$. Results of \cite{fulling-sweeny-wald, ford-parker} then guarantee that the state remains well defined at any other time. 

We will conclude this section with a few comments on  $|0_{\eta_0}\rangle$. First, what is the level of restriction imposed by conditions  (\ref{cond1}) and (\ref{cond2})? Does the desired instantaneous vacuum exist in the FLRW solutions that are most commonly used in cosmology, or only for very specific forms of the scale factor $a(\eta)$?  As a first exercise it is interesting to examine the situation in Minkowski space-time, in which $a(\eta)$ is a constant. In that case the adiabatic subtraction terms become $C_{\rho}(\eta_0,k,m)=w(\eta_0)=\sqrt{k^2+ m^2}$ and $C_{\rho}(\eta_0,k,m)=k^2/(3 w(\eta_0))$, and equations  (\ref{omega}) and (\ref{v}) give $\Omega_k(\eta_0)=w(\eta_0)$ and $V_k(\eta_0)=0$. Thus the solution exists and defines precisely the standard Minkowski vacuum, just as one would hope.

\begin{figure}[h]
\begin{center}
$\begin{array}{ccc}
\includegraphics[width=0.48\textwidth,angle=0]{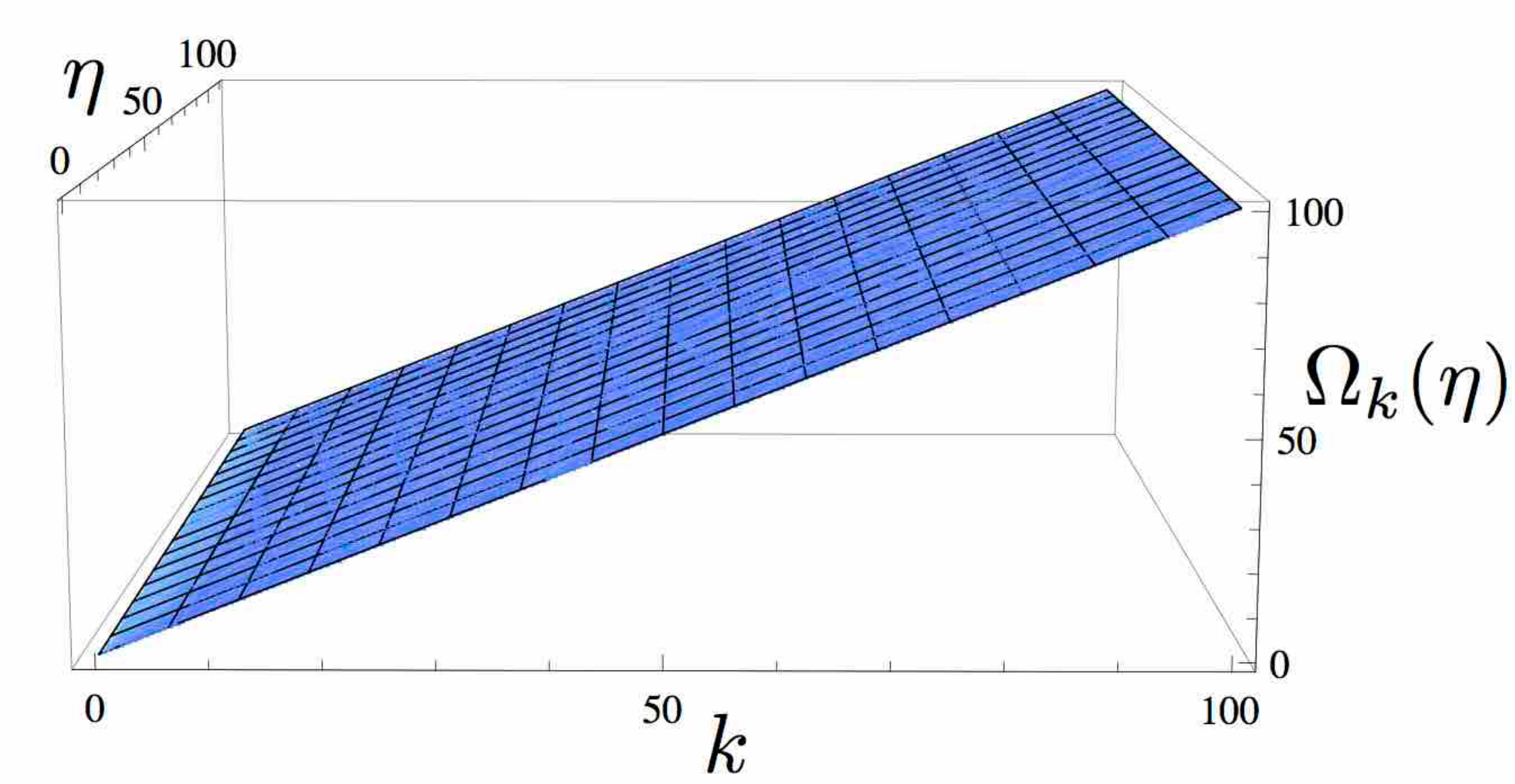} 
&
\includegraphics[width=0.48\textwidth,angle=0]{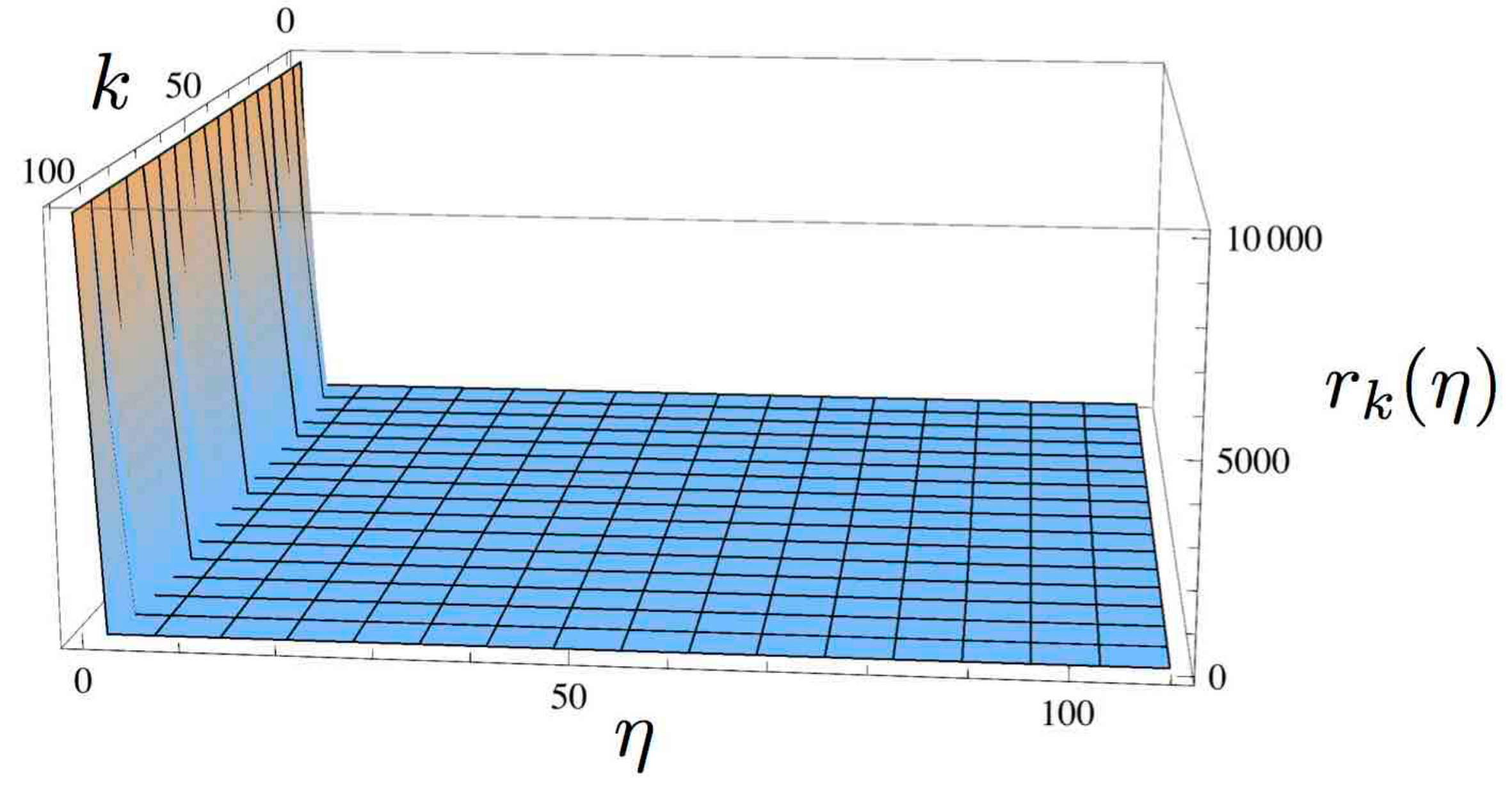}  
\end{array}$
\caption{FLRW space-time sourced by radiation: The left panel shows the behavior of $\Omega_k(\eta)$, defined in Eq. (\ref{omega}) as a function of $\eta$ and $k$ and the right panel, the behavior of $r_k(\eta)$ defined in Eq. (\ref{cond2}). Here we have set $m = 10^{-5}$ in the natural Planck units. In this model the scale factor has the form $a(\eta) = a_{0} \eta$ whence the singularity occurs at $\eta=0.$ For the range of parameters considered, both functions remain finite and positive except very near the big bang, satisfying the necessary and sufficient condition (\ref{cond1}) and (\ref{cond2}) of the existence of the instantaneous vacuum at time $\eta$.
\label{fig:f1}}
\end{center}
\end{figure}

\begin{figure}[h]
\begin{center}
$\begin{array}{ccc}
\includegraphics[width=0.48\textwidth,angle=0]{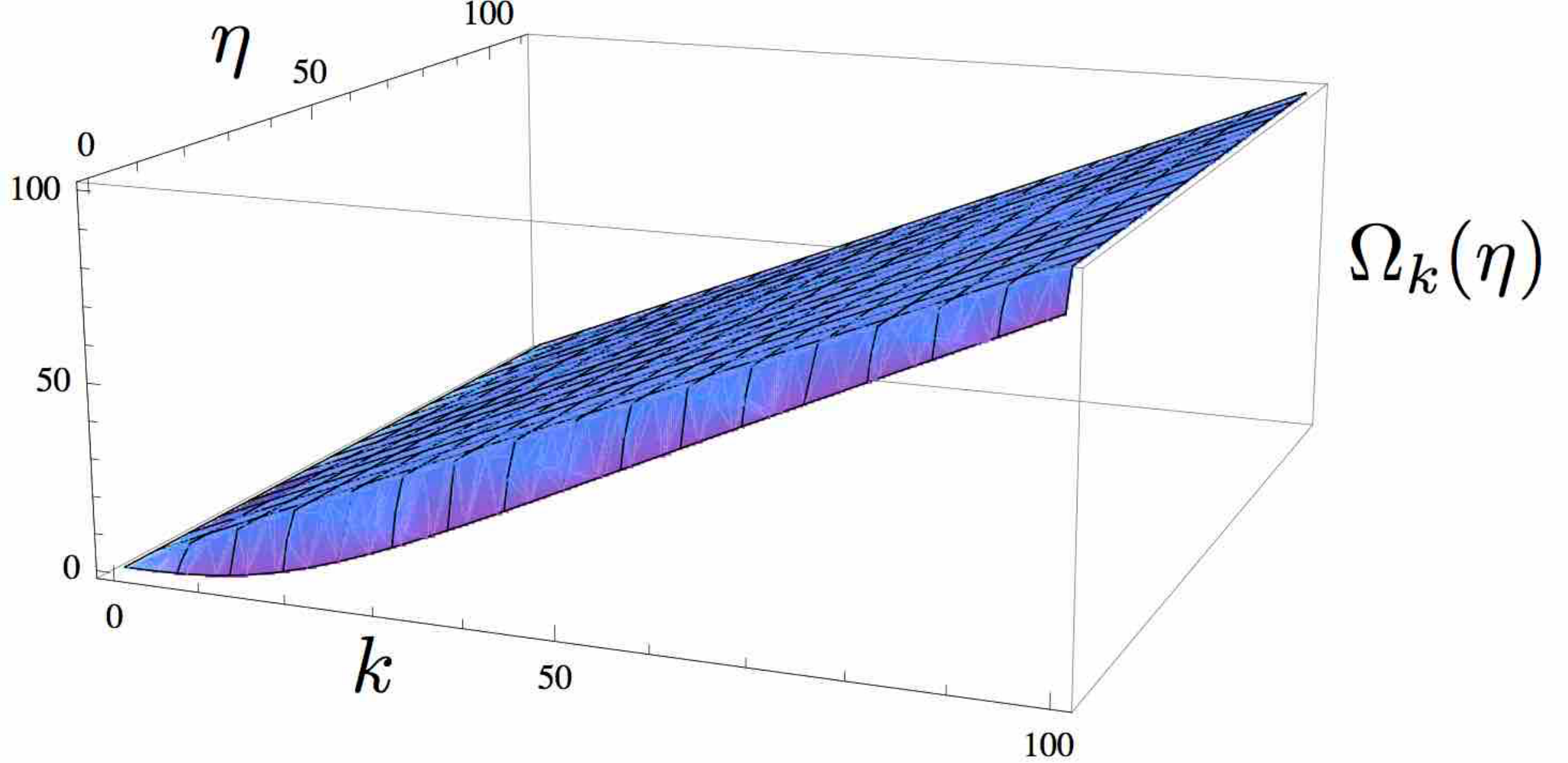} 
&
\includegraphics[width=0.48\textwidth,angle=0]{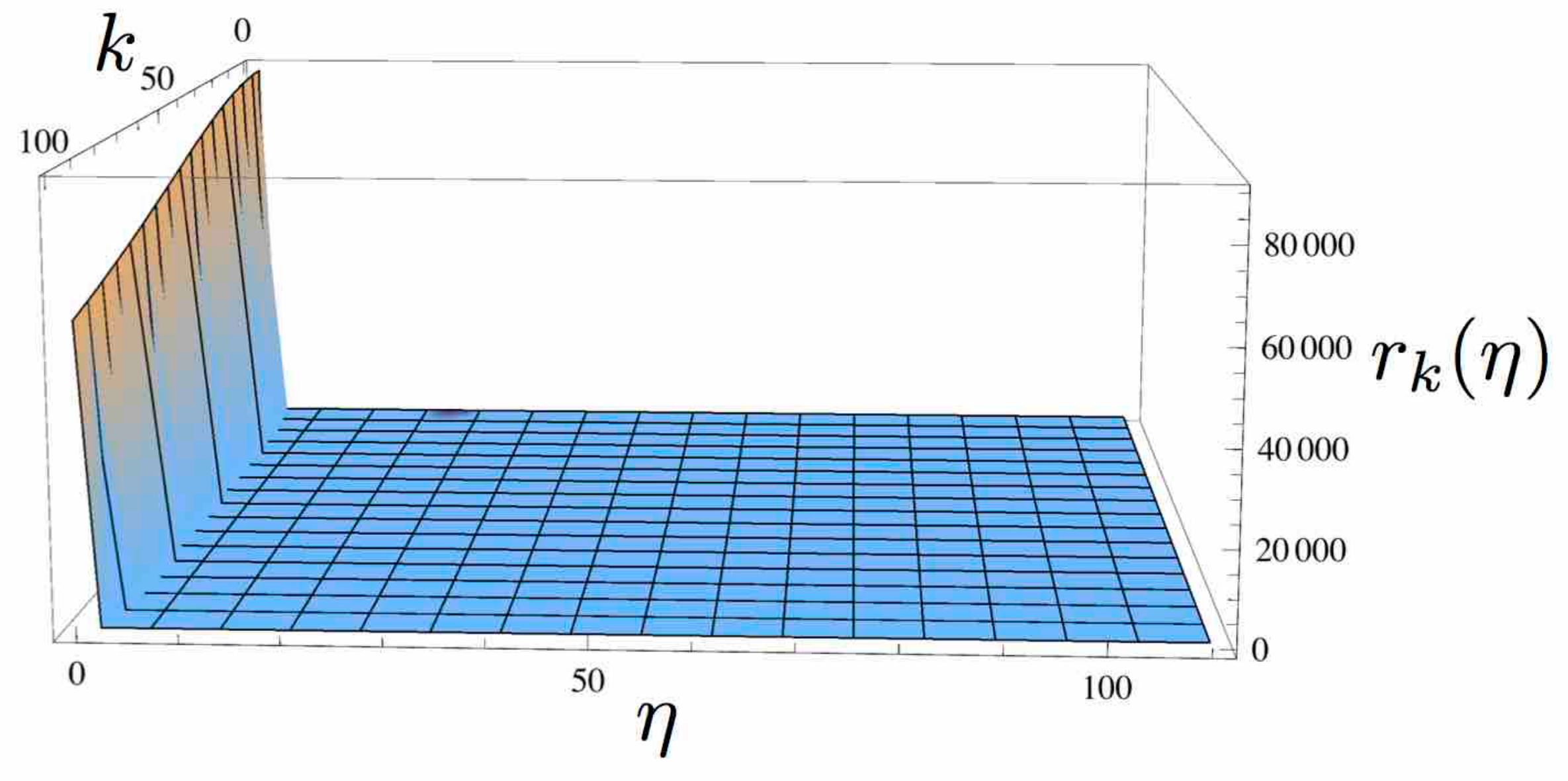}  
\end{array}$
\caption{FLRW space-time sourced by nonrelativistic matter: As in Fig 1, the left panel shows the behavior of $\Omega_k(\eta)$ (defined in (\ref{omega})) and the right panel, the behavior of $r_k(\eta)$ defined in (\ref{cond2}). The mass parameter is again $m = 10^{-5}$ in the natural Planck units. Again, except very near the big bang singularity ($\eta=0$), both functions remain finite and positive. Thus for the range of $\eta,k$ shown, the existence of the instantaneous vacuum $|0_{\eta_{0}}\rangle$ is ensured.
\label{fig:f2}}
\end{center}
\end{figure}

\begin{figure}[h]
\begin{center}
$\begin{array}{ccc}
\includegraphics[width=0.48\textwidth,angle=0]{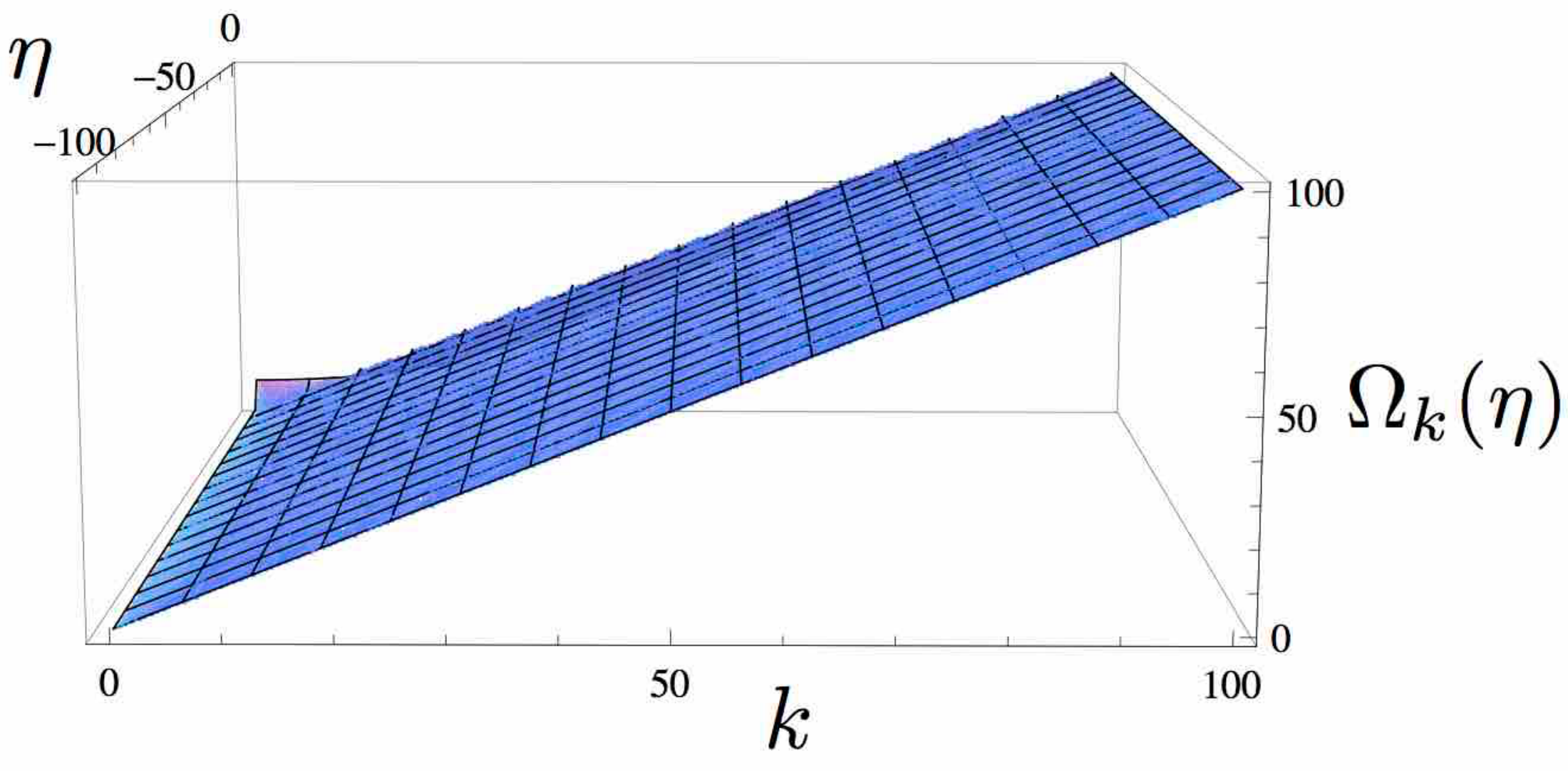} 
&
\includegraphics[width=0.48\textwidth,angle=0]{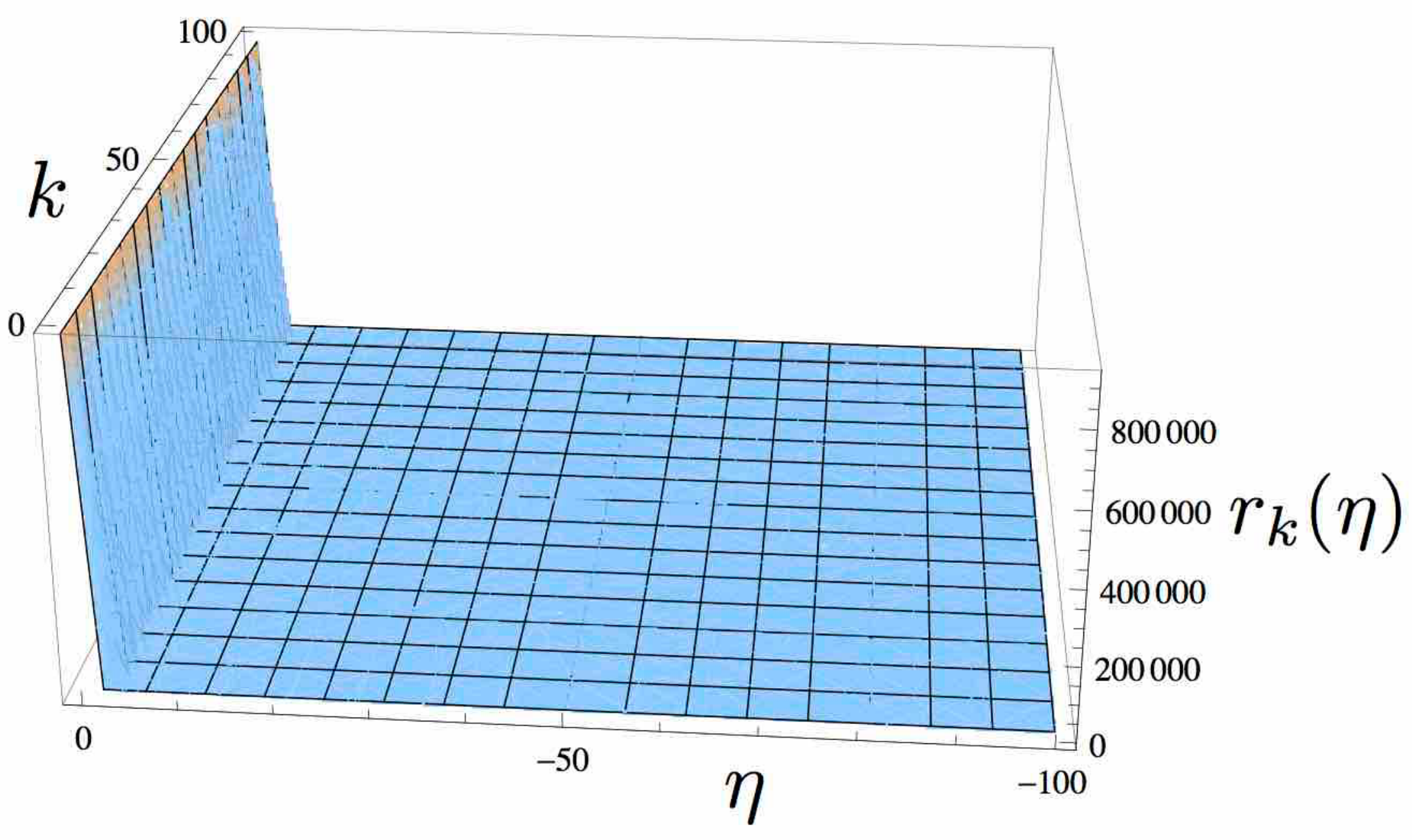}  
\end{array}$
\caption{De Sitter space-time: Again, the left panel shows the behavior of $\Omega_k(\eta)$  and the right panel, the behavior of $r_k(\eta)$. In this plot we have set $m= H = 10^{-5}$ in the natural Planck units. Now, $\eta=0$ corresponds to future infinity which is not part of space-time. For the values of $\eta, k$ considered, the functions remain finite and positive, ensuring the existence of the instantaneous vacuum $|0_{\eta_{0}}\rangle$.
\label{fig:f3}}
\end{center}
\end{figure}

 For time-dependent scale factors $a(\eta)$ we have  checked numerically that the necessary and sufficient conditions (\ref{cond1}) and (\ref{cond2}) are satisfied in the following cosmological space-times: i) a radiation dominated FLRW universe in which the scale factor has the form $a(\eta)=a_0\  \eta$; ii) a matter dominated FLRW universe, for which  $a(\eta)=a_0\,  \eta^{2}$; and de Sitter space-time with $a(\eta)= -1/(H \eta)$ and $H$ constant. A number of numerical simulations were carried out. Except at and very near the big bang, we found no values of $m$, $H$, $k$ and $\eta$ at which conditions (\ref{cond1}) and (\ref{cond2}) are not satisfied. We include three illustrative plots (FIG 1 -- FIG 3) of the behavior of $\Omega_{k}(\eta)$ and $r_{k}(\eta)$ (defined in (\ref{cond2})), one for each of these space-times, and for $m = H = 10^{-5}$ in the natural Planck units with $c=G_{\rm N}= \hbar =1$. These plots show that both functions remain finite and positive for all plotted range of $k$ and $\eta$, as it is required by (\ref{cond1}) and (\ref{cond2}). Their qualitative behavior can be understood as follows. $\Omega_k(\eta)$ behaves like $\Omega_k(\eta)\sim k+ [{\hbox{\rm (terms with two derivatives of $a$)/$k$ + higher adiabatic order terms]}} $ for large values of $k$ compared to the mass or the curvature. Therefore, except for very small values of $k$, we have $\Omega_k(\eta) \sim k$. The three plots exhibit this $\eta$-independence and linear growth in $k$ of $\Omega_{k}(\eta)$. Next, consider $r_{k}(\eta)$. It behaves like $r_{k}(\eta) \sim  a^{\prime}/a+\mathcal{O}(k^{-2})$. For values of $k$ that are large compared to the mass or the curvature, the term $a^{\prime}/a$ dominates and the plots are approximately $k$-independent. But for small $k$, there is $k$-dependence. Furthermore since $a^{\prime}/a \sim \eta^{-2}$ in all three cases considered, there is a strong growth when approaching  $\eta=0$. This growth may seem to be `abrupt' in the first three figures. But that is an artifact of the very large scale used in the vertical axis showing values of $r_{k}(\eta)$. In FIG 4, which zooms in at small values of $r_{k}(\eta)$, one sees that the growth is gradual, following the $1/\eta^{2}$ behavior.

Our simulations showed that larger values of the $\eta$ and $k$, and other choices of $m$ and $H$ did not alter the final conclusions: The state $|0_{\eta}\rangle$ continued to exist. However, since the search was done numerically, it could not be exhaustive. If one is interested in using the instantaneous vacuum in a specific situation, one has to use the values of $\eta_{0}$ and $m$ (and $H$) of interest and verify that conditions (\ref{cond1}) and (\ref{cond2}) are satisfied.

\begin{figure}[h]
\begin{center}
$\begin{array}{ccc}
\includegraphics[width=0.48\textwidth,angle=0]{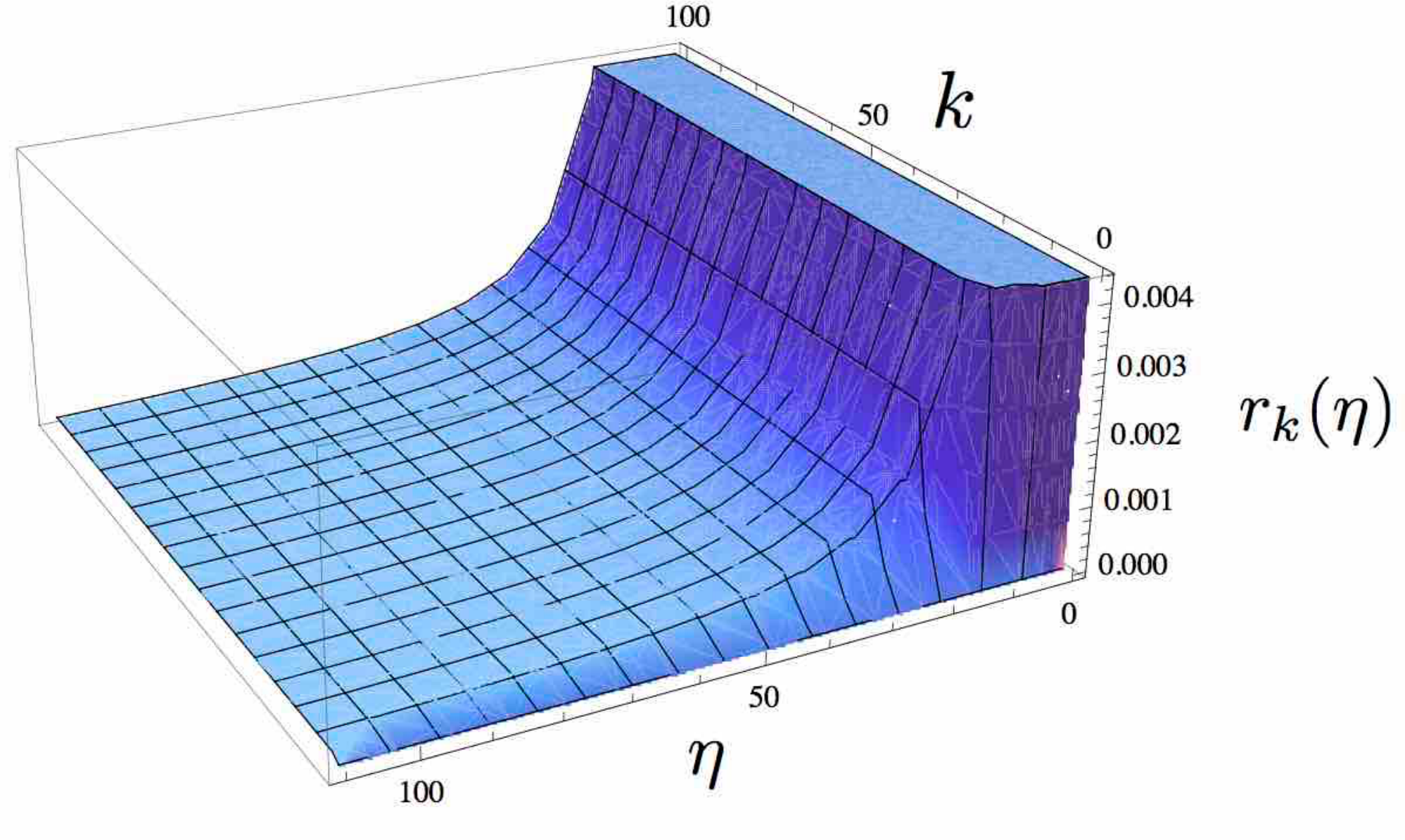} 
&
\includegraphics[width=0.48\textwidth,angle=0]{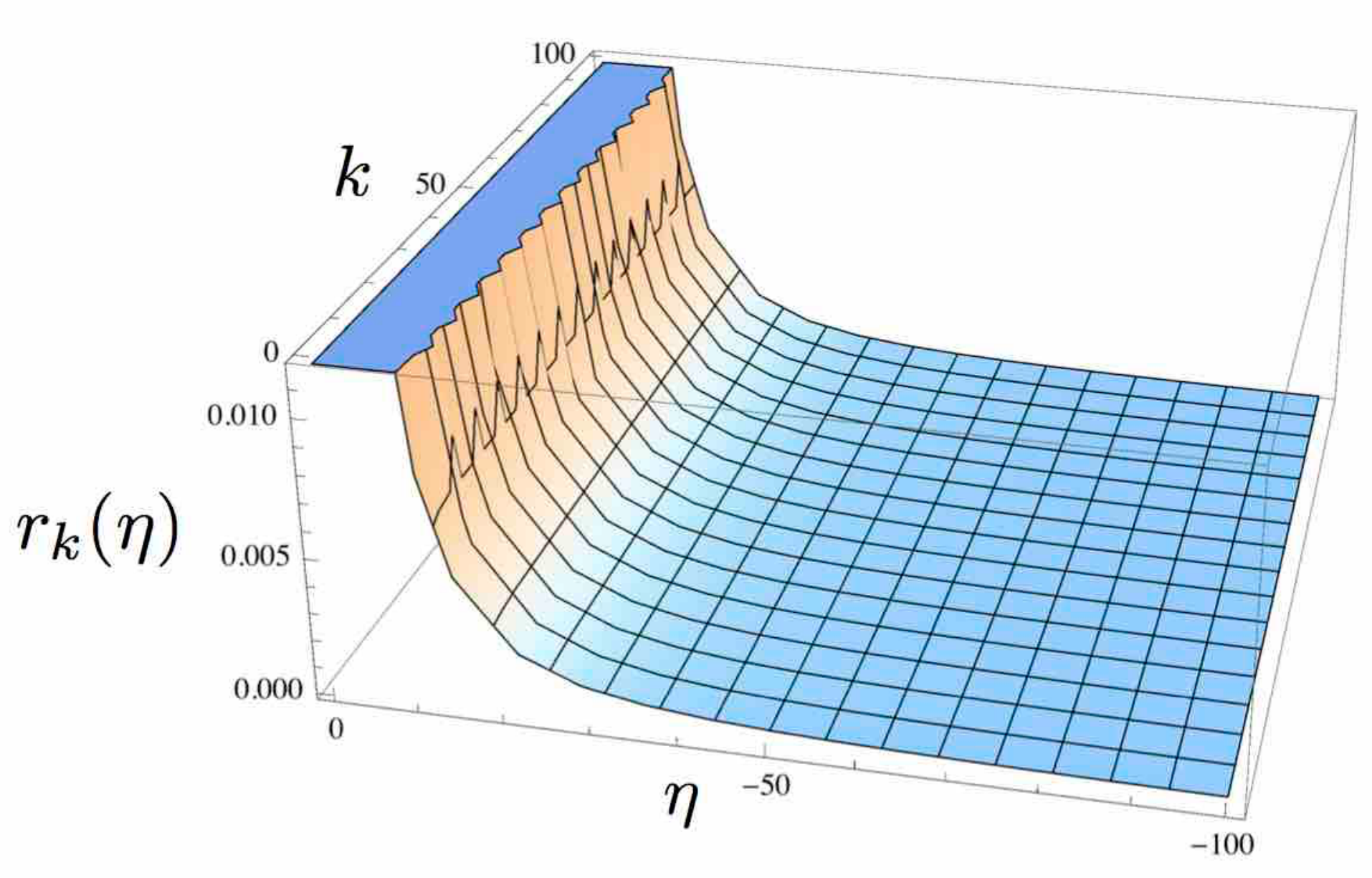}  
\end{array}$
\caption{Growth of $r_{k}(\eta)$ in de Sitter (left) and radiation filled FLRW (right) space-times: This figure zooms-in on the region in which $r_{k}(\eta)$ starts growing as one approaches $\eta=0$. The growth goes as $a^{\prime}/a \sim 1/\eta^{2}$. Behavior is similar for the matter-filled FLRW universe.
\label{fig:f4}}
\end{center}
\end{figure}

For those space-times for which the group of isometries is larger than the Euclidian group (homogeneity and isotropy), one would not expect the preferred instantaneous vacuum to automatically agree with states singled out by the full symmetry group. This is because $|0_{\eta_0}\rangle$ is constructed using the preferred cosmological foliation: the Euclidean group is tied to this foliation, and the local geometry  used in our construction---the scale factor and its first four time derivatives---also refers to this foliation. It is not required to be invariant under any additional symmetries. For instance, in de Sitter space-time the cosmological foliation is not preserved by the full isometry group and our instantaneous vacuum does not agree with the Bunch-Davies vacuum which \emph{is} invariant under the full de Sitter group. 

It is not difficult to show that although $|0_{\eta_0}\rangle$ has vanishing expectation value for the energy-momentum tensor at time $\eta_0$, it is {\em not} an eigenstate of the operator $\hat T_{ab}$ or the energy operator $\int \dd^3x  \, a^3 \hat \rho$ at that time, unless $a(\eta)$ is very special. On the other hand, it is also not difficult to see that if $|n_{\vec{k}}\rangle$ is an eigenstate of the number operator $N^{(\eta_0)}_{\vec{k}}=(A_{\vec{k}}^{(\eta_0)})^{ \dagger}A^{(\eta_0)}_{\vec{k}}$ with eigenvalue $n_{\vec{k}}$, where $A^{(\eta_0)}_{\vec{k}}$ are the annihilation operators associated with $|0_{\eta_0}\rangle$, then the total energy  in the state $|n_{k}\rangle$ at time $\eta_0$ is given by $a(\eta_0)^3 (2\pi)^{-3} \big(\rho[\varphi_k(\eta_0)] - C_{\rho}(\eta_{0}, k, m)\big)\,  \times n_{\vec{k}}$. Therefore,  $\big(\rho[\varphi_k(\eta_0)] - C_{\rho}(\eta_{0}, k, m)\big)$ can  be interpreted as the average energy density at $\eta_0$ per quantum (of $N_{\vec{k}}^{(\eta_{0})}$) in a comoving volume in position as well as momentum space.

\section{Minimally coupled massless scalar field} \label{s4}

The massless limit requires special attention because of the potential infrared divergences. In the adiabatic approach, a renormalization energy scale $\mu>0$ needs to be specified for massless fields to handle these divergences \cite{anderson-eaker}. The introduction of this scale does not add  further ambiguity to the renormalization procedure. This is because different choices of the scale $\mu$ translate to the addition of a term proportional to the geometric tensor $H^{(1)}_{ab}$  (defined in expression (\ref{H})  in the Appendix)) to the renormalized energy-momentum tensor, and the ambiguity of adding such a term is already present in the axiomatic renormalization approach \cite{Waldbook}. For massless fields this is in fact all the freedom one has in the choice of renormalization scheme, since $H^{(1)}_{ab}$ is the only conserved, geometric tensor with the same dimensions as $T_{ab}$ available in FLRW. Therefore, one can think of the choice of the scale $\mu$ as encoding all the freedom in the choice of the renormalization scheme.

As in the massive case, the preferred instantaneous vacuum at time $\eta_0$ is characterized  by two real parameters $\Omega_{k}(\eta_0)$ and $V_{k}(\eta_0)$ that solve the equations 
\be \rho[\varphi_k(\eta_0)] =C^{(m=0)}_{\rho}(\eta_0,k,\mu) \, ; \hspace{1cm} p[\varphi_k(\eta_0)] =C^{(m=0)}_{p}(\eta_0,k,\mu) \, , \ee
where $\rho[\varphi_k(\eta_0)]$ and $p[\varphi_k(\eta_0)] $ are given by (\ref{rho1}) and (\ref{p1}) with $m=0$.  The massless subtraction terms $ C^{(m=0)}_{\rho}(\eta_0,k,\mu)$ and $ C^{(m=0)}_{p}(\eta_0,k,\mu)$ are given by expressions (\ref{Crhom0}) and (\ref{Cpm0}) with $\xi=0$. Solutions to these equations are
\be \label{omegam0} \Omega_{k}(\eta_0)=-\frac{ k^2}{3 \, a(\eta_0)^4\,(  C^{(m=0)}_p(\eta_0,k,\mu)-   C^{(m=0)}_{\rho}(\eta_0,k,\mu))} \, ; \ee 
\be \label{vm0} V^{(\pm)}_{k}(\eta_0)=2\frac{a'(\eta_0)}{a(\eta_0)}\mp 2 \sqrt{-k^2+4\, a(\eta_0)^4 \,  C^{(m=0)}_{\rho}(\eta_0,k,\mu)\, \Omega_{k}(\eta_0)-\Omega_{k}^2(\eta_0)} \, .\ee
Therefore, the preferred instantaneous state exists whenever $\Omega_{k}(\eta_0)$ is positive and the radicand of (\ref{vm0}) is non-negative
\be \label{condm01} \infty  > \Omega_{k}(\eta_0)>0\, ,\ee
\be \label{condm02} -k^2+4\, a(\eta_0)^4\,   C^{(m=0)}_{\rho}(\eta_0,k,\mu)\, \Omega_{k}(\eta_0)-\Omega_{k}^2(\eta_0)\ge 0 \, .\ee
 As in the massive case, $(\Omega_{k}(\eta_0)$,\,\, $V^{(+)}_{k}(\eta_0))$ provide a 4th adiabatic order vacuum for $a'(\eta_0)\ge 0$ and  $( \Omega_{k}(\eta_0)$ and $V^{(-)}_{k}(\eta_0)$, for  $a'(\eta_0)\le 0$. One can check that the above conditions  (\ref{condm01}) and  (\ref{condm02})  are indeed satisfied in the most common FLRW background used in cosmology. In the constant $a(\eta)$  limit the resulting state agrees with the Minkowski vacuum.
  
However, the vacuum state we have just found  is not satisfactory: the two-point function diverges in the infrared limit, $k\to 0$, whence the state fails to satisfy our regularity requirement. The situation is similar to the well-known problem of the massless limit of the Bunch-Davies vacuum in de Sitter space \cite{allen}, and the solution is the same as in that case \cite{allen-folacci}. It suffices to change the zero mode, $\varphi_{k=0}$, to bypass the problem. For the massless Bunch-Davies vacuum, the resulting quantum state is no longer exactly de Sitter invariant. However, the deviation from de Sitter invariance appears in a single mode with  $k=0$.  As a consequence, this state is considered to be physically interesting and is widely used in the context of inflation. In our case, the result of modifying the prescription (\ref{omegam0}) and (\ref{vm0}) for $k=0$ yields a state in which energy momentum at $\eta_0$ fails to vanish, but only due to the contributions of modes with arbitrarily small $k$.

\section{Conformally coupled massless scalar field}\label{s5}

In this section we discuss the subtleties that arise in the conformally coupled case, with $m=0$ and $\xi=1/6$. The classical expression for the energy-momentum tensor takes the form

\be T_{ab}=\frac{2}{3} \nabla_a\phi\nabla_b\phi-\frac{1}{6}g_{ab}\, g^{cd} \nabla_c\phi \nabla_d\phi -\frac{1}{3} \phi\nabla_a\nabla_b\phi+\frac{1}{12} g_{ab}\Box \phi+\frac{1}{6} (R_{ab}-\frac{1}{4}R\, g_{ab})\phi^2 \, ,\ee
where $R_{ab}$ is the Ricci tensor and $R$ its trace.  This tensor is manifestly traceless, $g^{ab}T_{ab}=0$. This vanishing trace is the source of the issue that we now describe.  

At the quantum level, the adiabatically renormalized VEV of $\hat T_{ab}$ in a homogeneous and isotropic state takes again the perfect fluid form (\ref{pf}), with

\be \label{rho6} \langle \rho \rangle_{\rm ren} = \frac{1}{(2\pi)^3} \int \dd^3k \, \left(\rho[\varphi_k]-C_{\rho}^{\rm (cnf)}(\eta,k)\right)\ee
\be\label{p6}  \langle p \rangle_{\rm ren}:= \frac{1}{(2\pi)^3} \int d^3k \, \left( p[\varphi_k] -C_{p}^{\rm (cnf)}(\eta,k)\right) \, ,\ee 
where $\rho[\varphi_k]=\frac{1}{2 a^2} (|\varphi_k'+\frac{a'}{a}\varphi_k|^2+k^2 |\varphi_k|^2)$ and $p[\varphi_k]= 1/3\, \rho[\varphi_k]$. The adiabatic subtraction terms for conformal coupling, $C_{\rho}^{\rm (cnf)}(\eta,k)$ and $C_{p}^{\rm (cnf)}(\eta,k)$,  can be found in the Appendix  (expressions (\ref{Crhom0}) and (\ref{Cpm0}) with $\xi=1/6$). The renormalized VEV of the trace is given by 
\bea \label{T} \langle T \rangle_{\rm ren}&=& 3\, \langle p \rangle_{\rm ren} -\langle \rho \rangle_{\rm ren}= - \frac{1}{(2\pi)^3} \int \dd^3k \, \left(3\, C_{p}^{\rm (cnf)}(\eta,k)-C_{\rho}^{\rm (cnf)}(\eta,k)\right)\nonumber   \\&=&\frac{1}{180 \, (4\pi)^2} \left(\Box R+ R_{ab}R^{ab}-\frac{1}{3}R^2\right)\, .\eea
Notice that, as a consequence of the vanishing classical trace, this VEV is {\em independent} of the mode functions $\varphi_k$, {\it i.e.}, independent of the vacuum in which the expectation value is evaluated: $\langle T \rangle_{\rm ren}$ arises entirely from renormalization subtractions. This is the well-known trace or conformal anomaly \cite{capper-duff,desser-duff-isham} (see also \cite{birrell-davies}). As a consequence, there is obviously no state for which the renormalized VEV of the energy-momentum tensor is zero. Therefore, the preferred instantaneous vacuum does not exist for a massless, conformally coupled scalar field in spatially flat FLRW space-time, unless $a(\eta)$ is very special, {\it e.g.} constant. Note also that this result is not a peculiarity of the adiabatic  approach; it extends to any other regularization scheme. This is because the value of the trace anomaly is nonzero in all renormalization procedures satisfying Wald's axioms \cite{Waldbook}. Only the coefficient multiplying $\Box R$ in  (\ref{T}) changes from one scheme to the other.
 
In the absence of a state which makes the VEV of all components of the energy-momentum tensor equal zero,  one could ask if there exist homogeneous and isotropic vacuum states for which one of the two independent components, the energy density {\em or} the pressure, has zero VEV  at a given time. The answer is also in the negative: for a generic $a(\eta)$, there is no vacuum of 4th adiabatic order with zero energy density or zero pressure at a given time for the conformally coupled scalar field.
 
\section{Discussion}\label{s6}

The problem of selecting preferred vacua for quantized fields in cosmological space-times is interesting not only  because of its conceptual importance, but also because the issue is directly relevant to the computation of primordial cosmic perturbations in the early universe. In these computations one needs to specify the quantum state for perturbations at some ``initial'' time $\eta_0$. In the inflationary scenario one uses the fact that the background is close to de Sitter space-time, and selects a vacuum by extending the Bunch-Davies vacuum state to quasi-de Sitter space-times (see e.g. \cite{Weinberg}).% 
\footnote{Note, however, that from a mathematical physics perspective, there is an infinite dimensional ambiguity in extending the notion to the \emph{near} de Sitter situations that feature in the slow-roll scenario.}
In loop quantum cosmology \cite{LQC}, matter-dominated bounces \cite{matter-bounce} and ekpyrotic cosmologies  \cite{ekpyrotic}, `initial conditions' are specified in a phase which is far removed from the slow-roll, de Sitter-like expansion. Can one still single out a preferred initial state at such initial instants? In the cosmological literature  it is common to choose ``Minkowski-like" initial data for modes to select the desired vacuum, namely $\varphi_k(\eta_0)=1/\sqrt{2 w(\eta_0)}$ and  $\varphi'_k(\eta_0)=-i w(\eta_0)/\sqrt{2 w(\eta_0)}$. Although the resulting state is homogeneous and isotropic by construction, from a physical perspective, it is not satisfactory because it is fails UV regularity: it is neither adiabatic nor Hadamard.  In particular, there is no known systematic procedure to renormalize the expectation value of the energy-momentum tensor in such states. In addition, even if one computes the {\it difference} in energy density between two states defined using these initial data at two different times, $\eta_0$ and $\eta_1$, one finds a divergent result for generic $a(\eta)$. Another avenue pursued in the early literature was to try to select a state that would be the ground state of the instantaneous Hamiltonian operator. At first this strategy seems attractive from a conceptual standpoint. Indeed, it leads to the standard vacuum state in Minkowski space-time. But, as explained in section \ref{s1}, in curved space-times it faces two difficulties: dependence on the choice of canonical variables used in the definition of the Hamiltonian and failure to be ultraviolet regular \cite{fulling78}. 
 
By contrast, the instantaneous vacuum introduced in this paper is free of these limitations. First, by construction, it is regular to 4th adiabatic order. Therefore its ultraviolet behavior is such that the expectation value of the stress-energy operator to be well defined. Second, the construction refers only to 4-dimensional fields; no choice of canonically conjugate variables is necessary. Furthermore, the input used in the construction is just the local geometry, namely the scale factor and its first four time derivatives. Finally, the construction can be carried out to completion in the most widely used FLRW models. Yet, the state $|0_{\eta_{0}}\rangle$ it selects has the same intuitive connotations as the `ground state of the instantaneous Hamiltonian' that was avidly sought in the older literature. In fact, it can be regarded as an instantaneous ground state in a stronger sense since not only do energy density and pressure vanish in this state, but they do so mode by mode.

From the perspective of semiclassical gravity, states $|0_{\eta_{0}}\rangle$ have an interesting property. We will now  make a detour to spell it out in some detail. Recall that in semiclassical general relativity, the space-time metric is classical, the matter fields are quantum, and the stress energy tensor in the classical Einstein's equation is replaced by the expectation value of the stress-energy tensor operator. Let us consider the following perturbative expansion. To the \emph{zeroth order}, we have a classical metric $\mathring{g}_{ab}$ coupled to classical matter which, for simplicity, we will take to be a Klein Gordon field $\mathring{\phi}$, satisfying
\be \mathring{\Box}\mathring{\phi} - m^{2}\mathring{\phi} = 0\quad {\rm and} \quad \mathring{G}_{ab} = 8\pi \G \mathring{T}_{ab} \ee
where, as the notation suggests, $\mathring{\Box}$ and $\mathring{G}_{ab}$ refer to $\mathring{g}_{ab}$ and $\mathring{T}_{ab}$ is the stress-energy tensor of $\mathring{\phi}$ (on the space-time with metric $\mathring{g}_{ab}$). Next, we have a quantum field $\hat{\phi}^{(1)}$ satisfying 
\be \mathring{\Box} \hat\phi^{(1)}\, - m^2 \hat\phi^{(1)}  = 0  \ee
which we regard as a \emph{first order} perturbation. We are interested in calculating the back reaction on the classical metric due to this perturbation. So we can expand the metric as 
$$g_{ab} = \mathring{g}_{ab} + g^{(1)}_{ab} + g^{(2)}_{ab} + \ldots $$
and solve Einstein's equations order by order. Since the perturbation is order 1 and the stress energy is quadratic in the field, as is usual in the analysis of back reaction, we will seek a truncation which is consistent up to second order. In this truncated expansion, the stress-energy tensor  is to be constructed using the matter field $\mathring\phi + \hat\phi^{(1)}$ (and the metric to the appropriate order). The right side of Einstein's equation will feature the expectation value of this operator in a quantum state. In the final argument we will use the instantaneous vacuum $|0_{\eta_{0}}\rangle$ but for now let us allow it to be a general vacuum $|0\rangle$ that satisfies only the symmetry and regularity conditions. Then, because $\langle 0|\nabla_{a}\hat\phi^{(1)}|0\rangle = 0$, it follows that the first order metric perturbation satisfies the \emph{homogeneous} equation:
\be \label{1storder} G_{ab}^{(1)}\, \equiv \, \mathring{\Theta}\, g^{(1)}_{ab} \,=0\, , \ee
where $\mathring{\Theta}$ is a second order differential operator constructed from the zeroth order metric $\mathring{g}_{ab}$.%
\footnote{That the equation is homogeneous is at first surprising. Had we worked in the classical theory, because $\nabla_{a}\phi^{(1)} \not=0$, $T^{(1)}_{ab}$ would not be zero and would act as a source for first-order scalar perturbations in the metric. Similarly, if the metric perturbations were operators $\hat{g}^{(1)}_{ab}$ ---as $\hat\phi^{(1)}$ is--- the right hand side would be the operator $\hat{T}^{(1)}_{ab}$ which, unlike its expectation value in the state $|0\rangle$, is nonzero. By contrast, in the semiclassical framework, since the metric is classical and the matter field is quantum, Einstein's equation necessarily involves expectation values, and the expectation value $\langle 0|\hat{T}^{(1)}_{ab}|0 \rangle$ vanishes. Thus, the fact that $g^{(1)}_{ab}$ satisfies a homogeneous equation is a peculiarity of (the perturbative expansion in) semiclassical gravity. For a more general discussion of unforeseen features, see section VI.D of \cite{aan2}.} 
Solutions to the homogeneous equation (\ref{1storder}) represent tensor modes. Since we are calculating only the back reaction on the metric created by the scalar field, we are led to choose the solution $g^{(1)}_{ab} =0$. 

Nontrivial back reaction appears at second order, via the second order Einstein's equation. Since $g^{(1)}_{ab} =0$, this equation reduces to
\be \label{2ndorder} G^{(2)}_{ab}\, \equiv\, \mathring{\Theta}\, g^{(2)}_{ab}\, =\, \langle 0|\hat{T}_{ab}|0\rangle \, , \ee
where the right side features the renormalized stress-energy tensor, which is quadratic in the first order perturbations. We can now perform an initial value formulation of this equation with $\eta=\eta_{0}$ as the initial instant. The resulting scalar and the vector constraints are inhomogeneous elliptic equations for the linearized 3-metric $q^{(2)}_{ab}$ and extrinsic curvature $k^{(2)}_{ab}$, with source terms 
$$\langle 0|\hat{T}_{ab}|0\rangle\,\mathring{n}^{a} \mathring{n}^{b} \quad {\rm and} \quad \langle 0|\hat{T}_{ab}|0\rangle\mathring{n}^{a} \mathring{q}^{bc}$$ 
where $\mathring{n}^{a}$ is the unit normal to and $\mathring{q}^{ab}$ the intrinsic metric on the surface $\eta =\eta_{0}$, defined by $\mathring{g}_{ab}$. Now, the main point is that if we were to use $|0_{\eta_{0}}\rangle$ in place of a generic vacuum $|0\rangle$, then the source terms on the right sides of these constraint equations vanish. Therefore the constraint equations on $q^{(2)}_{ab}$ and $k^{(2)}_{ab}$ \emph{become homogeneous} at time $\eta=\eta_{0}$. Therefore, their solutions provide initial data for transverse traceless modes in $g^{(2)}_{ab}$. Again, because we are interested in the back reaction only due to the scalar field $\hat{\phi}^{(1)}$, \emph{we are led to choose the solution with} $q^{(2)}_{ab} (\eta=\eta_{0}) =0$ and $k^{(2)}_{ab} (\eta=\eta_{0}) =0$. This choice provides a natural way to eliminate the freedom to add a solution of the homogeneous equation to  solutions of (\ref{2ndorder}), which is necessary, in any case, to select the physically appropriate solution representing the back reaction only due to $\hat{\phi}^{(1)}$.  In this scheme, the pair $q^{(2)}_{ab} (\eta),\, k^{(2)}_{ab} (\eta)$ captures the leading order modifications to the background geometry at time $\eta$, because of the back reaction due to the quantum perturbation $\hat{\phi}^{(1)}$. This correction vanishes identically at $\eta=\eta_{0}$. \emph{In this precise sense, the state $|0_{\eta}\rangle$ has the property that the back reaction on geometry vanishes at $\eta=\eta_{0}$.}  This is interpretation of $|0_{\eta_{0}}\rangle$ within semiclassical gravity we wanted to spell out. Under evolution, the data $q^{(2)}_{ab} (\eta),\, k^{(2)}_{ab} (\eta)$ will be necessarily  nonzero because  $\langle 0_{\eta}|\hat{T}_{ab}|0_{\eta}\rangle$ is nonzero for $\eta\not= \eta_{0}$. Thus, if the scalar field $\hat{\phi}$ is in the state $|0_{\eta_{0}}\rangle$, to second order in perturbation theory there is a nontrivial back reaction on the geometry at any time $\eta \not=\eta_{0}$.

Because the back reaction vanishes at $\eta=\eta_{0}$, the state  $|0_{\eta}\rangle$  can be thought of as the analog of the standard vacuum in Minkowski space-time, albeit only at a given instant of time. This preferred instantaneous vacuum has been used in the study of cosmological perturbation in loop quantum cosmology, where initial conditions are specified at or near the bounce time \cite{aan1,aan3}. We expect it will be also useful in other scenarios to select `initial conditions' for cosmological perturbations. 

While $|0_{\eta}\rangle$ has several attractive features, as pointed out in section  \ref{s1}, our construction has an important caveat. We will conclude by reemphasizing this point. In quantum field theory in curved space-times, {\em a priori}, there is freedom to add certain local curvature terms to the expression of the renormalized stress-energy tensor \cite{Waldbook}. In any given renormalization scheme one obtains a specific expression; the freedom disappears. But different schemes can yield different renormalized  stress tensors. The defining property, $\langle 0_{\eta_{0}}| \hat{T}_{ab}|0_{\eta_{0}}\rangle = 0$, of our preferred instantaneous vacuum $|0_{\eta}\rangle$ refers to the adiabatic scheme, where one carries out a mode by mode subtraction. In the FLRW models, the adiabatic scheme gives the same results as DeWitt-Schwinger point-splitting regularization \cite{birrell, navarro-salas-del-rio}. But another scheme could well lead to a different preferred instantaneous vacuum. This is the caveat. The nontrivial feature of the construction is the \emph{existence} of a consistent scheme to select a preferred instantaneous state which succeeds in bypassing the limitations of other procedures, and which can be used in the most common cosmological models. Moreover, the fundamental equation of semiclassical gravity, $G_{ab} = 8\pi \G\, \langle \hat{T}_{ab} \rangle$, is meaningful only within a specific renormalization scheme that is used to give meaning to the right-hand side. Therefore, in any case, a choice has to be made to analyze issues such as the back reaction. Our construction uses a scheme that is well tailored for FLRW space-times and therefore widely employed in the cosmological literature. 

\section*{Acknowledgments}
We have benefited from discussions with and comments from A. Ladha, N. Morris, J. Navarro-Salas,  J. Pullin, and E. Wilson-Ewing.
This work was supported by the NSF grants PHY-1403943 and PHY-1205388, the Eberly research funds of Penn State, and the Marie Curie Fellowship program of the EU.

\begin{appendix}
\section{Adiabatic subtraction terms}
\label{a1}

In this Appendix we spell out the adiabatic subtraction terms for the energy density and pressure first for the massive, minimally coupled scalar field, and then for the massless field (see also \cite{anderson-parker}):

\bea \label{Crhoxi0}  C_{\rho}(\eta,k,m)&=&\frac{w}{2a^4}+\frac{(2wa'+aw')^2}{16a^6w^3}+\frac{1}{256 a^7w^7}(-24aw^2a'w'^2-120a^2wa'w'^3-45a^3w'^4+ \nonumber\\&+ &64w^4a'^2a''+112a w^3a'w'a''+16a^2w^2w'^2a''+16aw^4a''^2+16 aw^3a'^2w''+ \nonumber\\&+&112a^2w^2a'w'w''+40a^3ww'^2w''+16a^2w^3a''w''+4a^3w^2w''^2-32w^4a'a'''- \nonumber \\&-&16a^2w^3w'a'''-16a^2w^3a'w'''-8a^3w^2w'w''')\eea

\bea \label{Cpxi0} C_{p}(\eta,k,m)&=&\frac{w^2-m^2a^2}{6a^4 w}+\frac{1}{48 a^6 w^5}(12w^4a'^2+12 a w^3a'w'+3m^2a^4w'^2+9a^2w^2w'^2\nonumber\\&- &4m^2 a^3w^2a''-8a w^4a''-2m^2a^4ww''-4a^2w^3w''+
\frac{1}{768 a^7 w^9}\Big[256 w^6a'^2a''\nonumber\\&-&8w^4(10w^2a''^2+a'^2(9w'^2-6w w'')+2 wa'(-31w'a''+10wa'''))\nonumber\\&+ &8m^2 a^4w^2(25w'a''-10ww'a'''+2w(-5a''w''+wa''''))+8a^2w^3(4m^2wa'^2a''\nonumber\\&+ &a'(-45w'^3+42ww'w''-6w^2w''')+2 w (28w'^2a''-7wa''w''-13ww'a'''\nonumber\\&+ &2w^2a''''))+m^2 a^5(-315w'^4+420 w w'^2w''-60w^2w''^2-80w^2w'w'''+8w^3w'''')+\nonumber\\&+ &
a^3w^2(-765w'^4+960ww'^2w''+8ww'(10m^2a'a''-23ww''')\nonumber\\&-&4w^2(16m^2a''^2+27w''^2+8m^2a'a'''-4ww''''))\Big]\eea

\bea  \label{Crhom0} C^{(m=0)}_{\rho}(\eta,k,\mu,\xi)&=&\frac{k}{2a^4}+(1-6\xi)\frac{a'^2}{4 a^6 k}+(1-6\xi)\frac{1}{(k^2+a^2 \mu^2)^{3/2}}\frac{1}{288a^2} H^{(1)}_{00}+\nonumber\\&+& \frac{\mu^2}{128a^2(k^2+a^2 \mu^2)^{11/2}}\Big[(16 k^6 (1-6\xi)-4k^4\mu^2 (-43+252\xi)a^2+\nonumber\\&+&24k^2\mu^4a^4(3-2\xi)+27\mu^6a'^4(-7+32\xi))+8aa'^2a''(k^2+\mu^2a^2) (12k^4(1-6\xi)^2\nonumber\\&+&6k^2\mu^2a^2(7-68\xi+144\xi^2)+\mu^4a^2(37-264\xi+432\xi^2))-\nonumber\\&-&4a^2a''^2(k^2+\mu^2a^2)^2(4k^2(-1+6\xi)+\mu^2a^2(-5+24\xi)) +\nonumber\\&+&8a^2a'a'''(k^2+\mu^2a^2)^2(4k^2(-1+6\xi)+\mu^2a^2(-5+24\xi))
\Big]\eea

\bea \label{Cpm0} C^{(m=0)}_{p}(\eta,k,\mu,\xi)&=&\frac{k}{6a^4}-(1-6\xi)\frac{-3a'^2+2 aa''}{12 k a^6}-(1-6\xi)^2\frac{1}{288 a^2(k^2+a^2\mu^2)^{3/2}}H^{(1)}_{xx}+\nonumber\\&+& \frac{\mu^2}{384a^2(k^2+a^2 \mu^2)^{13/2}}\Big[3a'^4(-16k^8(1-6\xi)+4 k^6\mu^2a^2 (-33+196\xi)+\nonumber\\&+&8k^4\mu^4a^4(-83+502 \xi)+k^2\mu^6a^6(-429+1024\xi)-72\mu^8a'^4(-7+32\xi))+\nonumber\\&+&4aa'^2a'' (k^2+\mu^2 a^2) (16 k^6(5-66\xi+216\xi^2)+4k^4\mu^2 a^2(59-972\xi+3672\xi^2)\nonumber\\&+&8k^2\mu^4a^4(85-996\xi+2376\xi^2)+\mu^6a^6(755-5136\xi+7776\xi^2))
-\nonumber\\&-&8a^2a'a'''(k^2+a^2 \mu^2)^2 (4k^4(7-78\xi+216\xi^2)+3k^2\mu^2a^2 (31-288\xi+576\xi^2)\nonumber\\&+&\mu^4a^4 (79-552\xi+864\xi^2))
-4a^2(k^2+\mu^2a^2)^2 (a''^2(8k^4(4-51\xi+162\xi^2)\nonumber\\&+&3k^2\mu^2 a^2 (39-392\xi+864\xi^2)+2\mu^4a^4 (53-384\xi+648\xi^2))+\nonumber\\&+&2 aa'''' (k^2+\mu^2a^2)(4k^2(-1+6\xi)+\mu a^2 (-5+24\xi))) \Big]\, ,\eea
where 
\be \label{H} H^{(1)}_{ab}=2 g_{ab} \Box R-2 \nabla_a\nabla_b R+2 R R_{ab}-\frac{1}{2}g_{ab}R^2 \, , \ee
and   $H^{(1)}_{00}=\eta^a\eta^bH^{(1)}_{ab}$ , $H^{(1)}_{xx}=x^ax^bH^{(1)}_{ab}$ are its time-time and $x$-$x$ components, respectively.

\end{appendix}

\end{document}